\documentclass[12pt]{iopart}

\usepackage{graphicx}
\usepackage{dcolumn}
\usepackage{bm}
\usepackage{subfigure}
\usepackage{color}      

\date{\today}

\begin{document}

\newcommand{\U}{U_{\mathrm{eff}}}
\newcommand{\dxsq}{$d_{x^2-y^2}$}
\newcommand{\dxz}{$d_{xz}$}
\newcommand{\dzsq}{$d_{z^2}$}
\newcommand{\dxy}{$d_{xy}$}
\newcommand{\dyz}{$d_{yz}$}

\newcommand{\dsplit}{\dxsq--\dxz}

\newcommand{\Edisp}{E_{\mathrm{disp}}}
\newcommand{\Edc}{E_{\mathrm{dc}}}
\newcommand{\clo}{VO$_4$Cl$_2$}

\def \mysize{0.6}
\newcommand{\mub}{\mu_{\mathrm{B}}}
\newcommand{\ang}{\AA$^3$/atom}

\title{The role of magnetic order in VOCl}

\author{M.\ Ekholm,$^{1,2}$ A.\ Sch\"{o}nleber,$^{1}$ S.\ van Smaalen$^{1}$}
\address{$^1$ Laboratory of Crystallography, University of Bayreuth, 95440 Bayreuth, Germany}
\address{$^2$ Link\"oping University, SE-581 83 Link\"oping, Sweden}
\ead{marcus.ekholm@liu.se}

\begin{abstract}
\noindent VOCl and other transition metal oxychlorides are candidate materials for next-generation rechargeable batteries.
We have investigated the influence of the underlying magnetic order on the crystallographic  and electronic structure by means of density functional theory.
Our study shows that antiferromagnetic ordering explains the observed low-temperature monoclinic distortion of the lattice, which leads to a decreased distance between antiferromagnetically coupled V-V nearest neighbors.
We also show that the existence of a local magnetic moment removes the previously suggested degeneracy of the occupied levels, in agreement with experiments.
To describe the electronic structure, it turns out crucial to take the correct magnetic ordering into account, especially at elevated temperature.
\end{abstract}

\submitto{\JPCM}
\maketitle

\section{Introduction}
The ever increasing need of efficient energy storage solutions is currently driving the search for economical and environmentally friendly options beyond the lithium-ion battery.
Prototypes relying on shuttling of Cl$^-$ ions have been demonstrated to match lithium-based counterparts in energy capacity, but can be constructed from more abundant elements \cite{zhao14}.
Nevertheless, a crucial task is to find suitable electrode materials, which must be structurally stable in the electrolyte while offering optimal discharge capacity.
For the cathode, several promising candidates have been proposed among the metal oxychlorides, such as BiOCl \cite{zhao13,chen17}, FeOCl \cite{zhao16,yu17} and VOCl \cite{gao16}.
In order to take full advantage of their electrochemical and structural properties it is important to understand the electronic structure, and experimental design may therefore be combined with theoretical calculations \cite{zhao16}.

\begin{figure}[hbt!]
\begin{center}
\subfigure[\label{fig:overview}]{
\includegraphics[scale=1.6]{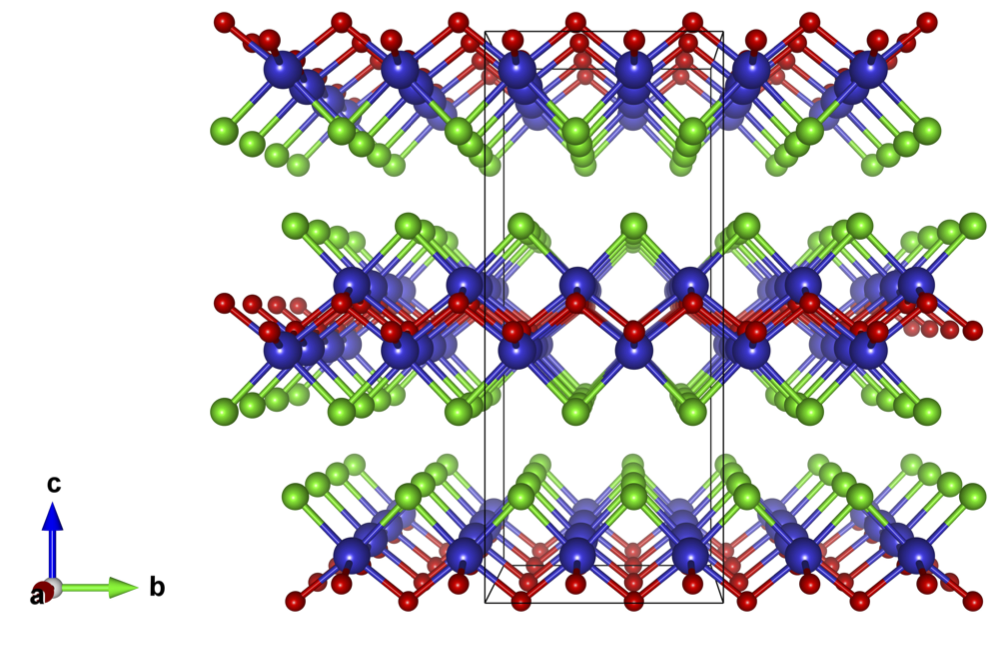}
}
\includegraphics[scale=0.07]{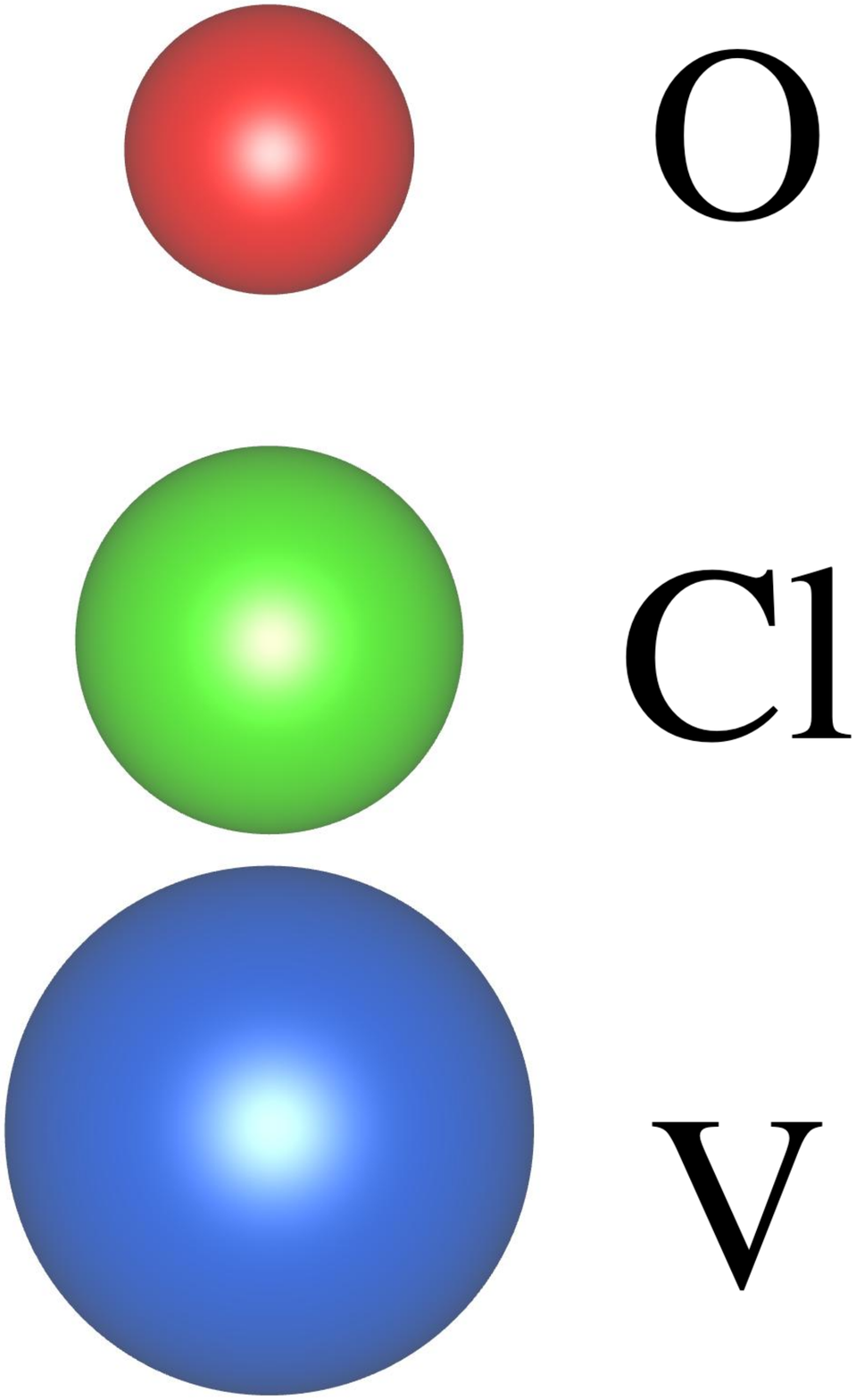}
\subfigure[\label{fig:octahedron}]{
\includegraphics[scale=0.5]{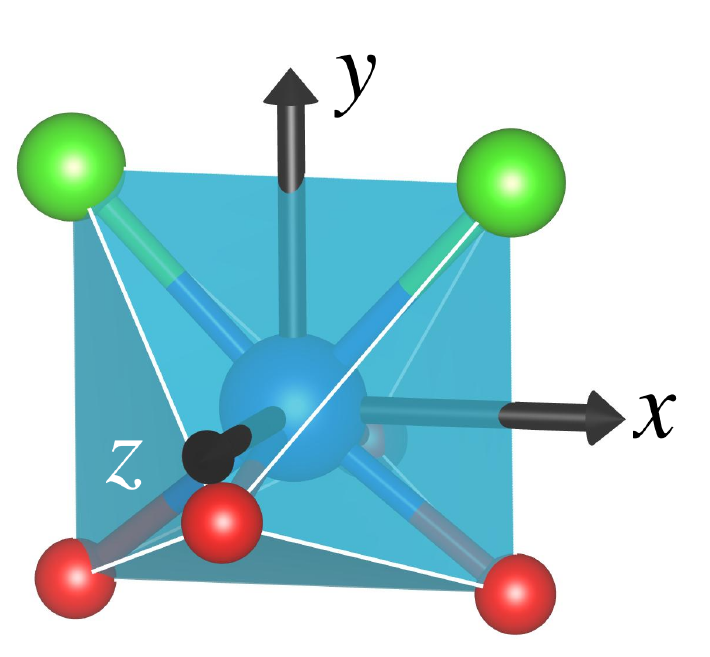}
}
\end{center}
\caption{
(a) Overview of the VOCl crystal structure.  
(b) Distorted VO$_4$Cl$_2$ octahedron with local coordinate frame $(x,y,z)$, with $x$ along $\mathbf{b}$, $y$ along $\mathbf{c}$ and $z$ along $\mathbf{a}$.
\label{fig:crystalstructure}
}
\end{figure}

However, transition metal compounds with a partially filled $3d$ shell show a highly non-trivial coupling between spin, orbital and lattice degrees of freedom \cite{imada98,dagotto05}, which still presents a significant challenge to solid-state theory.
In addition, the orthorhombic $M$OCl crystal structure consists of layers of interconnected, distorted $M$O$_4$Cl$_2$ octahedra separated by a van der Waals gap (see \fref{fig:crystalstructure}).
This may lead to highly anisotropic interactions due to reduced dimensionality.

Indeed, in the isostructural Mott insulator TiOCl, the Ti$^{3+}$ ions, which are in $d^1$ configuration, have been concluded to form quasi one-dimensional spin chains along $\mathbf{a}$ \cite{seidel03,sahadasgupta04,rotundu18}.
Upon cooling, orthorhombic TiOCl undergoes two successive phase transitions.
At 90 K the lattice becomes incommensurately modulated, with a small $\mathbf{c}$-axis monoclinic distortion of $0.023^\circ$ of the $\gamma$ angle \cite{schonleber08}.
At 67 K the distortion switches to $\mathbf{a}$-axis monoclinic,
while the Ti chains undergo a spin-Peierls
dimerisation \cite{fausti07,schonleber08}.
The strongly correlated nature of the electrons has been revealed in theoretical studies based on sophisticated many-body techniques going beyond the static mean-field description of $d$-$d$ interactions, \cite{sahadasgupta05} and also including non-local interactions to describe the electronic structure \cite{sahadasgupta07,aichhorn09}.

\begin{figure}[hbt!]
\begin{center}
\includegraphics[scale=0.1]{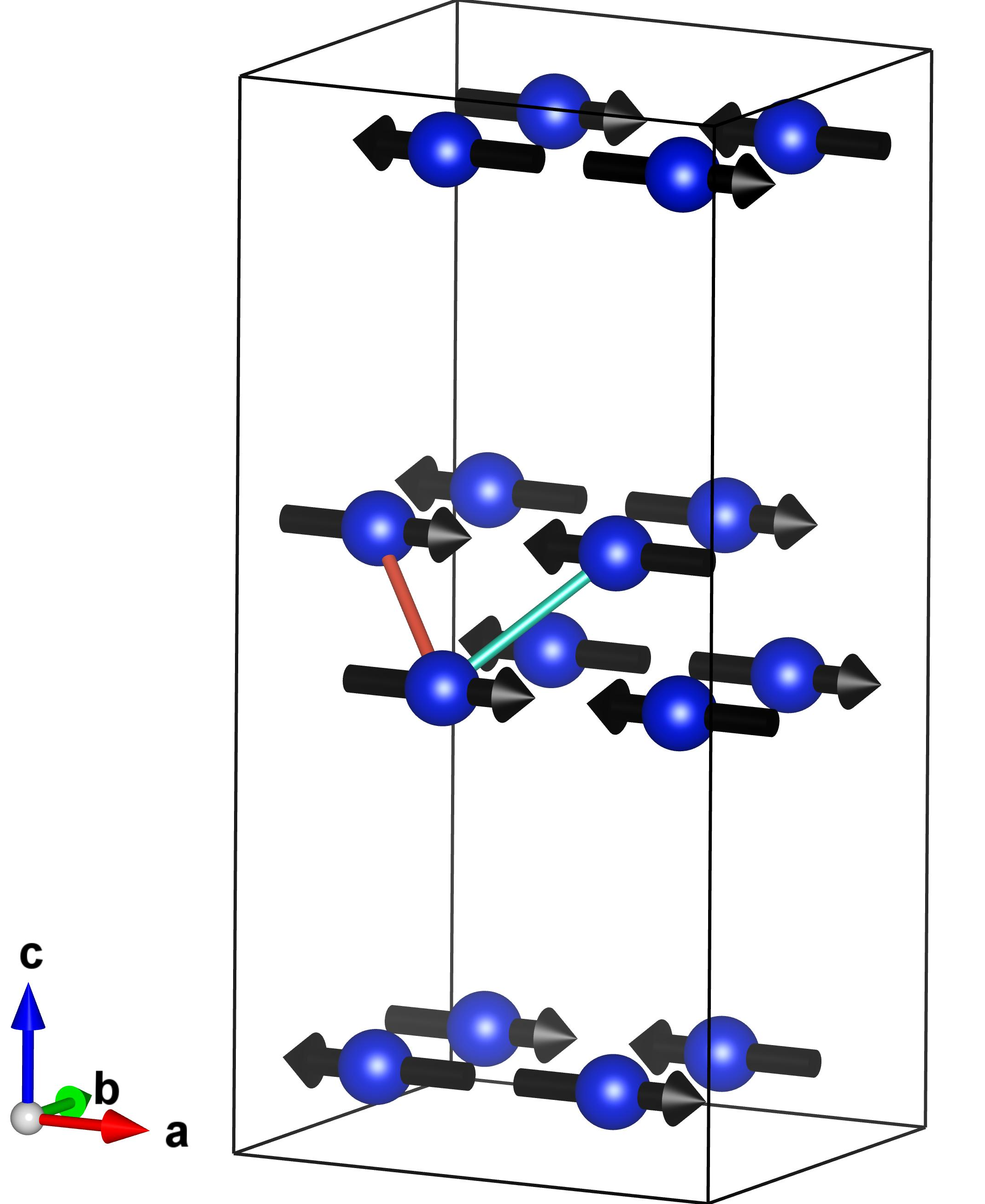}
\caption{ 
V-sublattice showing AFM-ordering. Nearest-neighbors along $0.5\mathbf{a}\pm 0.5\mathbf{b} + 0.1\mathbf{c} $ are indicated.
\label{fig:afm_order}
}
\end{center}
\end{figure}

VOCl, which features V$^{3+}$ ions in $d^2$ configuration, is much less studied.
Optical absorption measurements have shown VOCl to be insulating, with a band gap of 1.5--2.0 eV \cite{venien79,benckiser08}.
Neutron diffraction and magnetic susceptibility measurements indicate a magnetically disordered state with finite local moments at room temperature.
At the N\'{e}el temperature, $T_N\approx 80$~K, the system adopts a two-fold antiferromagnetic (AFM) superstructure  \cite{wiedenmann83, komarek09, glawion09}.
Two independent studies \cite{schonleber09,komarek09} have concluded the AFM order shown in \fref{fig:afm_order}.
The ordering transition at $T_N$ is accompanied by a monoclinic distortion, which preserves the overall symmetry of the lattice.
At 2~K, the monoclinic angle between the $\mathbf{a}$ and $\mathbf{b}$ axes deviates from orthogonality by $0.2^\circ$.
This distortion makes the distance between V$^{3+}$ ions with anti-parallel magnetic moments shorter by $\sim 0.01$~\AA.
Monoclinic distortions have also been observed in CrOCl and FeOCl \cite{schonleber09,angelkort09, zhang12}.

Theoretical studies have emphasized the importance of strong correlations in reproducing the electronic spectrum \cite{glawion09,bogdanov11}.
Glawion \etal \cite{glawion09} used density functional theory (DFT) \cite{hohenberg64,kohn65} and showed that the Perdew-Burke-Ernzerhof (PBE) generalized gradient approximation (GGA) \cite{perdew96} does not reproduce the insulating state.
However, the insulating gap correctly opened with the DFT+U approach \cite{anisimov91}, which suggests that the main effects of strong correlation are properly captured.
Nevertheless, the occupied \dxsq\ and \dxz\ levels were reported as nearly degenerate, in contrast to optical conductivity measurements, which indicate a splitting of 0.1 eV.
On the other hand, the splitting could be reproduced with explicit on an embedded ferromagnetic VOCl cluster in a study by Bogdanov \etal \cite{bogdanov11} which suggested that DFT does not capture the splitting.

In this work we show that PBE+U calculations correctly reproduce the splitting, and that the orbital ordering critically depends on the assumed magnetic order. 
We also perform full structural relaxation with different magnetic configurations to show that the monoclinic distortion is a direct consequence of AFM ordering.
Details of the calculations are given in \sref{sec:details}.
Our results are organized into the magnetically ordered phase (\sref{sec:results_monoclinic}) and the disordered phase (\sref{sec:results_orthorhombic}).
For the ordered phase we first show results for the structure within various approximations schemes while assuming AFM order.
Finally, we discuss our results in \sref{sec:conclusions}.

\section{Computational details}\label{sec:details}
We have used the projector-augmented waves (PAW) \cite{blochl94,kresse99} method as implemented in the Vienna ab-initio simulation package (VASP), version 5.4 \cite{vasp1,vasp2,vasp3}.
The plane wave energy  cut-off was set to 500 eV, and the V $3p$-electrons were treated as valence states.

AFM-ordered VOCl was modelled with a supercell constructed from $2\times 2\times 2$ unit cells with $Pmmn$ symmetry as given in reference \cite{komarek09}.
To sample the Brillouin zone we set up a $10\times 11 \times 5$ k-point grid and used the Monkhorst-Pack scheme \cite{monkhorst76}.
We have used the PBE \cite{perdew96} and PBE+U functionals.
The latter was based on the so-called Dudarev parametrization \cite{dudarev97}, which makes use of the spherically averaged elements of the screened Coulomb electron-electron interaction. 
The correction term is in this case parametrized by the so-called effective U-parameter $\U=U-J$, where $U$ and $J$ are the on-site Coulomb and Hund's rule coupling parameters, respectively.
Structural optimization was done by converging total energy to within 2.1 $\mu$eV / atom.
In order to treat the van der Waals interaction between the layers during structural relaxation, we used the D3 term with Becke-Johnson damping \cite{grimme10,grimme11}.
The equation of state (EOS) was determined by fitting calculated total energy versus volume to the Birch-Murnaghan equation \cite{birch47}.

For the paramagnetic state we have used a supercell of 108 atoms.
V atoms of spin up and down were distributed according to the special quasirandom structure (SQS) method  \cite{zunger90}.
The short-range order parameter was required to vanish for the first nine coordination shells.
A $6\times 7 \times 4$ k-point grid and the Monkhorst-Pack scheme was used.

We have calculated the orbital energies, $\epsilon _n$, as the first order moment:
\begin{equation}\label{eq:energylevel}
\epsilon _n =  \frac{\int g_n(E) E \,  \mathrm{d}E}{\int g_n(E) \, \mathrm{d}E}
\end{equation}
where $g_n(E)$ is the partial density of states (DOS) as a function of energy, $E$, and $n \in \left\lbrace \textrm{  \dxsq, \dxz, \dyz, \dzsq, \dxy} \right\rbrace$.

\section{Results}\label{sec:results}
\subsection{Low temperature phase}\label{sec:results_monoclinic}
\subsubsection{The monoclinic distortion}\label{sec:monoclinic}
\begin{figure}
\begin{center}
\includegraphics[scale=0.65]{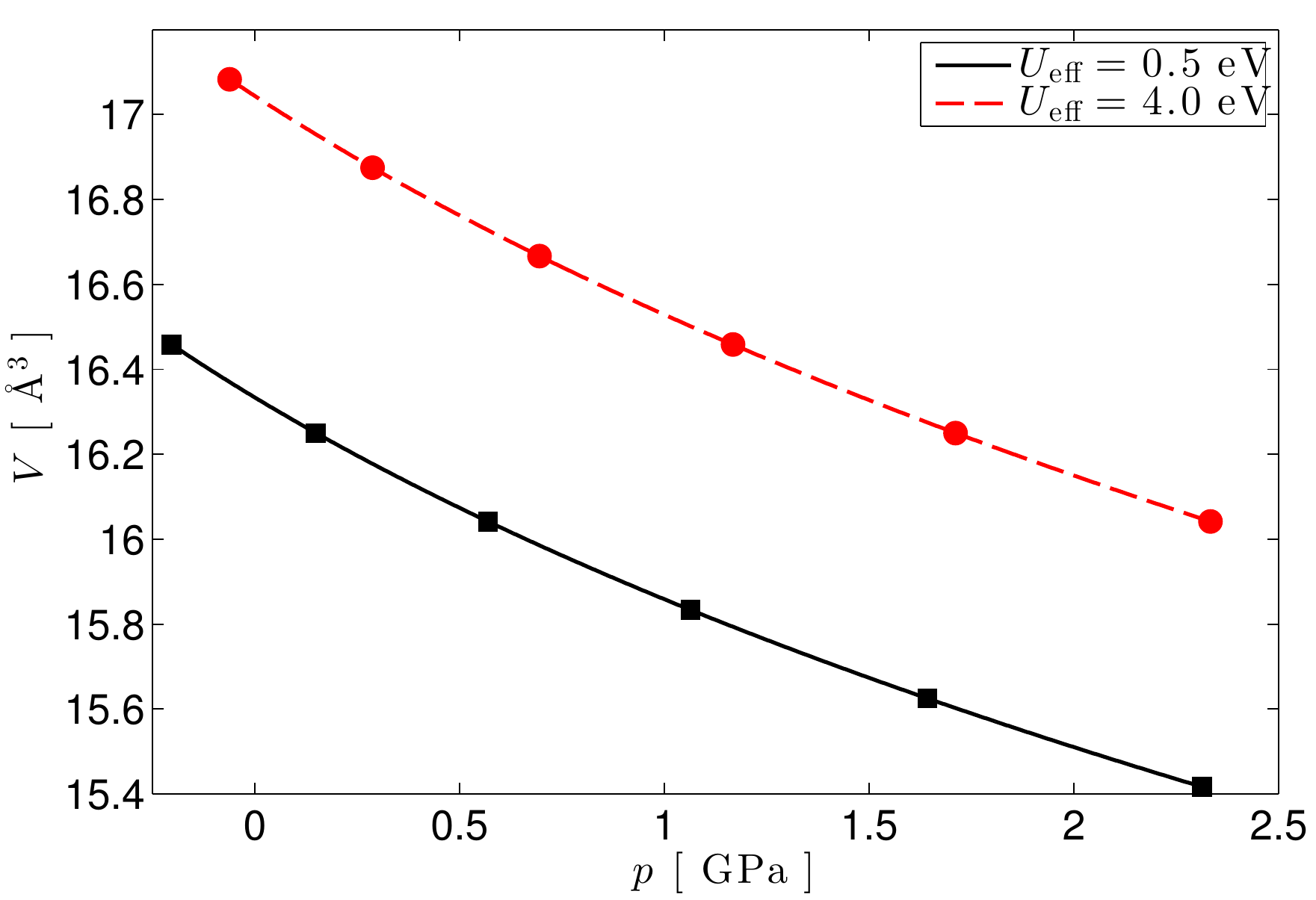}
\caption{Atomic volume as a function of pressure, obtained from fitting to the Birch-Murnaghan EOS.
\label{fig:VP}
}
\end{center}
\end{figure}

In order to investigate the relation between structure and magnetic order below the N\'eel temperature, we have used $2\times2\times 2$ supercell with AFM order among the V atoms, as illustrated in \fref{fig:afm_order}.
Relaxing the ion positions with the PBE+U+D3 approximation, we obtain the EOS parameters presented in \tref{tab:eos} for different values of the $\U$-parameter.
The corresponding volume-pressure curves are shown in \fref{fig:VP}.
\begin{table}[hbt!]
\caption{\label{tab:eos}
EOS and lattice parameters obtained by full structural relaxation for AFM-order with various approximations to the exchange-correlation functional.
 }
\begin{indented}
\item[]\begin{tabular}{c l c c c c c}
\br
               &                           & Expt.$^{\rm a}$ &  $\U=0.5$ eV & $\U=2$ eV &  $\U=4$ eV  \\
              \mr
  $V_0$ & [\AA$^3$/ atom]      & 16.30     & 16.3      & 16.7       & 17.0  \\ 
  $B_0$  &   [GPa]              &     -         &      28.1         & 26.8       & 27.8       \\
  $B'$     &                              &     -         &    13.1        & 13.6      & 11.8    \\
  $ b/a $     &                       &  0.873   &       0.86         & 0.87   &  0.87     \\ 
  $c/a $        &                     &  2.09      &      2.09         & 2.08   &  2.07    \\
  $\gamma$ &  [$^\circ$]  &  90.212 &   90.75    & 90.46 &   90.28     \\
  \br
  \end{tabular}
   \item[] $^{\rm a}$   {References\ \cite{schonleber09} (3.2 K) and\  \cite{komarek09} (2 K)}
  \end{indented}
\end{table}
With $\U=0.5$ eV, the unit cell volume is well reproduced, and the lattice parameter ratios are in good agreement with experiment. 
However, the monoclinic angle is larger than the 2~K experimental value of $90.2^\circ$.
Increasing the $\U$ term, the equilibrium volume increases while the monoclinic angle is decreased.
For $\U=4.0$ eV, we obtain $\gamma=90.28^\circ$, closer to the experimental value, but the unit cell volume is overestimated by approximately 4\%.
Nevertheless, the diagonal V-V distances (indicated in \fref{fig:afm_order}) become somewhat closer (0.02 \AA) to the experimental values with $\U=2$~eV than with $\U=4$~eV.

We find that the local spin magnetic moment for a V atom is $1.87\mub$, while the orbital  moment is $-0.074\mub$ with $\U=2$ eV, i.e., directed antiparallel to the spin moment.
Interestingly, experiments have reported a magnetic moment of ($1.48\pm 0.18$)$\mub$ \cite{wiedenmann83}, and more recently 1.3$\mub$ \cite{schonleber09,komarek09}, which is significantly lower than what may be expected from a system in $d^2$ configuration.
The exaggeration of the calculated magnetic moment is most likely due to the GGA+U static mean-field approximation.

With the assumed magnetic configuration (\fref{fig:afm_order}), the result $\gamma > 90^\circ$ for the monoclinic angle means that the distance between anti-parallel magnetic moments is smaller than between parallel moments.
Including spin-orbit coupling, we have calculated the magnetocrystalline anisotropy energies with respect to the crystallographic [100], [010], and [001] axes.
The anisotropy energy obtained with $\U=2$ eV is $E_{010}-E_{100} = 0.07$ meV / atom, and $E_{001}-E_{100} = 0.14$~meV / atom.
This indicates magnetic ordering along $\mathbf{a}$.
Our results regarding the magnetic ordering are thus in line with the interpretation of the diffraction data made in references \cite{komarek09} and \cite{schonleber09}, as well as previous single crystal magnetic susceptibility measurements by Wiedenmann \etal \cite{wiedenmann83}, which also indicated an ordered moment along $\mathbf{a}$.

\subsubsection{Electronic structure}\label{sec:elstruct}

\begin{figure}
\begin{center}
\subfigure[\label{fig:afm_u2_total}
]{
\includegraphics[scale=\mysize]{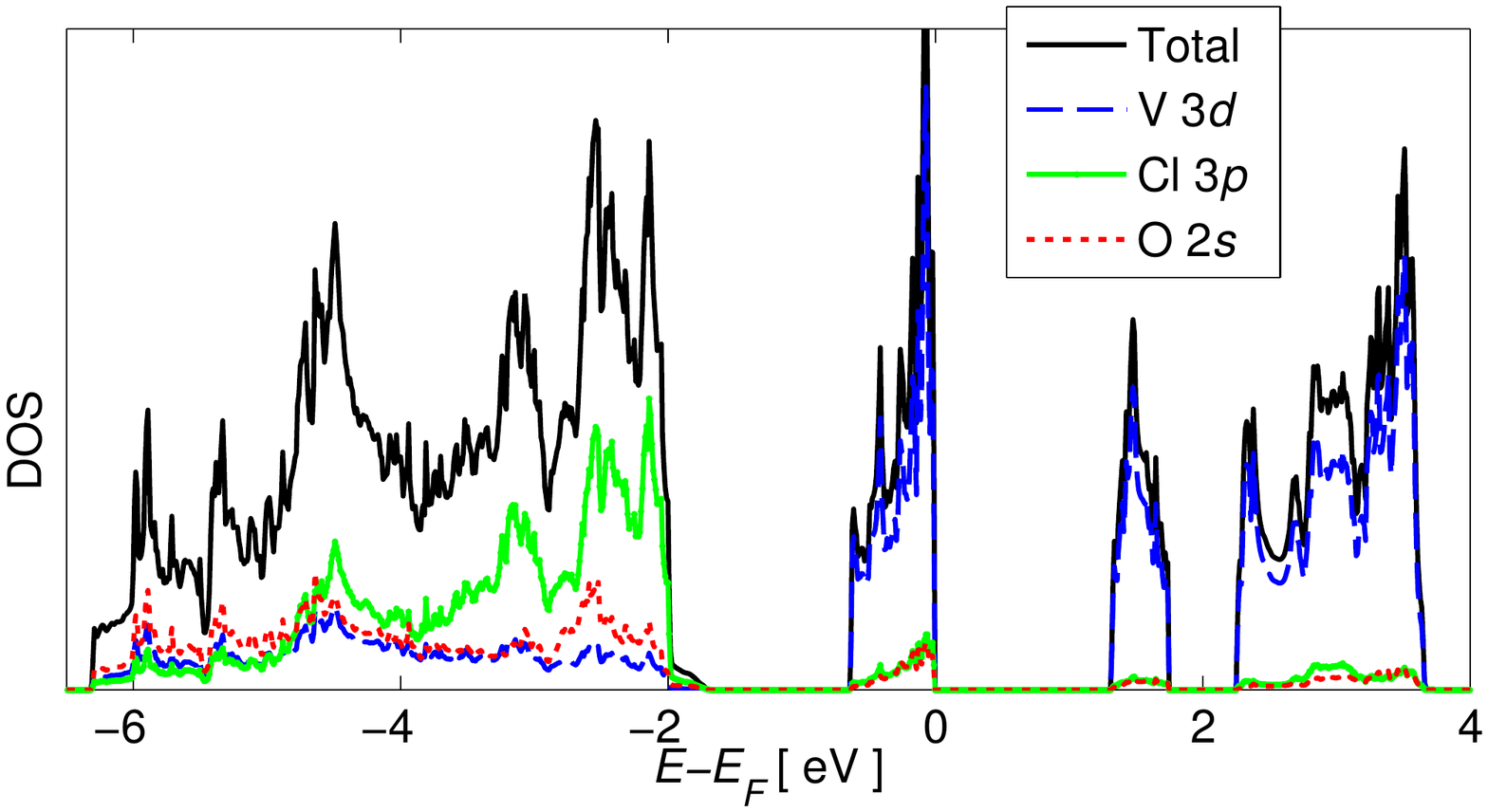}
}
\subfigure[\label{fig:afm_u2_partial}
]{
\includegraphics[scale=\mysize]{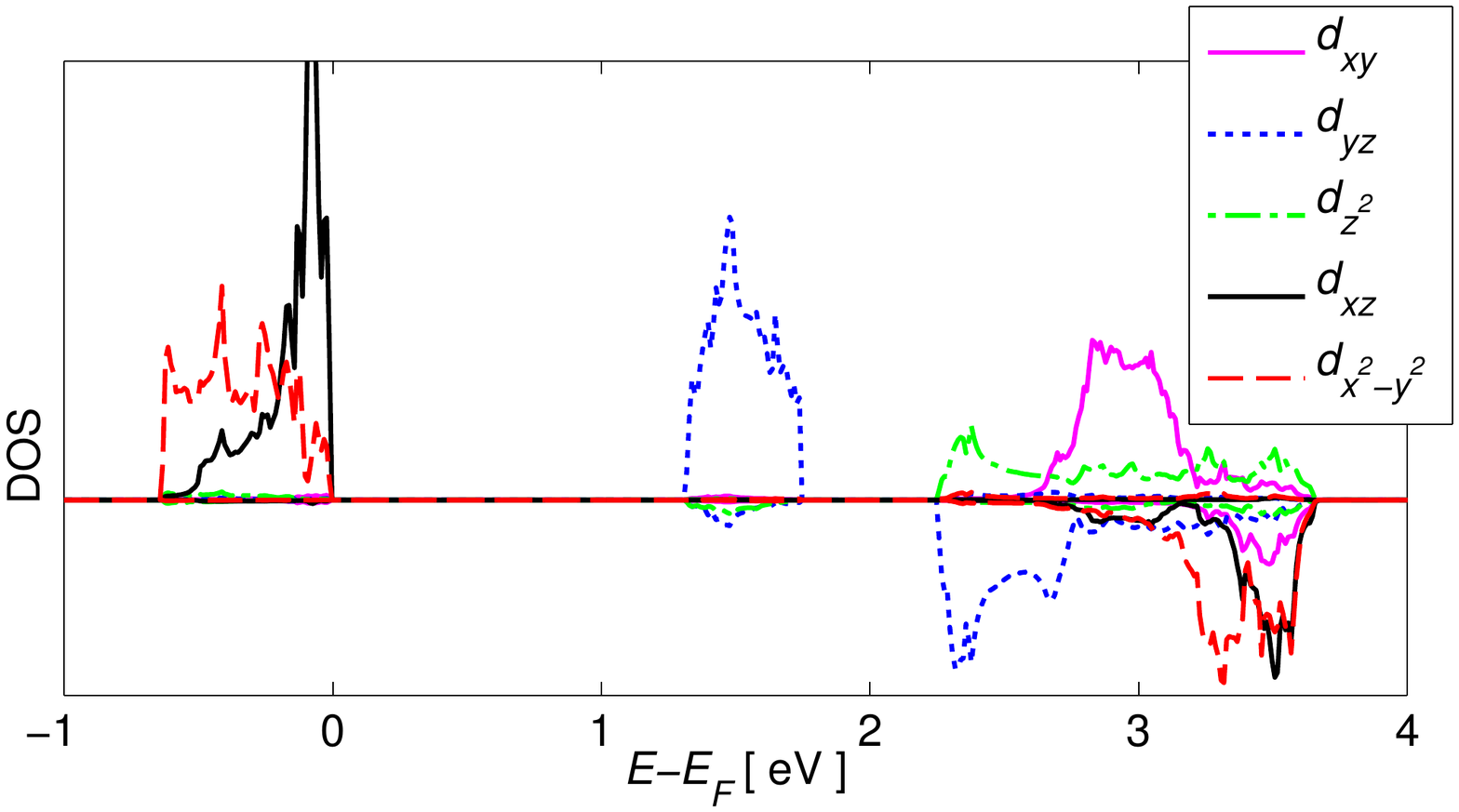}
}
\caption{DOS calculated for AFM-ordered VOCl using the PBE+U functional at the optimized lattice constants.
\label{fig:afm_u2_relaxed}
}
\end{center}
\end{figure}
In \fref{fig:afm_u2_total} we show the total DOS calculated with $\U=2$~eV for the optimized structure, assuming AFM order.
The electronic spectrum is seen to be divided in a low-binding energy part consisting of V $3d$-states, which is separated by a gap of 1.0~eV from a high-binding energy part of O $2p$-, Cl $3p$-states V $3d$-states.
This is consistent with the photoemission measurements of Glawion \etal, which seem to indicate a gap of $\sim 1 $ eV between the low- and high-binding energy parts.
 
An insulating charge gap of 1.3~eV separates the occupied and unoccupied V $3d$ states.
The width of both gaps will depend directly on the assumed $\U$ value, as pointed out in reference \cite{glawion09}.
A large value of $\U$ will open the insulating V-V gap but close the V-O/Cl gap.
We conclude that $\U=2$ eV is a reasonable choice to simultaneously reproduce crystal structure and electronic structure of antiferromagnetic VOCl.

In \fref{fig:afm_u2_partial} we show the site- and spin-projected DOS for a V-atom, in terms of the irreducible representation.
The local $(x,y,z)$ coordinate system is chosen so that $x=b$, $y=c$ and $z=a$ for the orthorhombic structure.
Orbital energies defined through equation \eref{eq:energylevel} are also summarized in \tref{tab:enorb} for different values of $\U$.
We find the occupied spin up \dxsq\ and \dxz\ states to be split by 0.19 eV.
This is contrast to reference \cite{glawion09}, which concluded these orbitals to be nearly degenerate based on DFT calculations.
However, Bogdanov \etal \cite{bogdanov11} found a splitting of about 0.1 eV with coupled cluster calculations, which agrees with optical conductivity measurements \cite{benckiser08}.
Since the correction of the U-term mainly affects the splitting between occupied and unoccupied levels, the \dsplit\ splitting is only weakly dependent on the $\U$-value.
The splitting is also insensitive to the monoclinic $\gamma$ angle, since the distortion does not change the shape of the octahedra, but only leads to a tilting.
It is interesting to compare with TiOCl, where resonant inelastic x-ray scattering (RIXS) indicate a clear splitting of 0.3--0.4 eV between the \dxsq\ and \dxz\ orbitals \cite{glawion11}.
For the unoccupied states, the DOS should not be compared to excitation energies. Nevertheless, we note that the order of the \dxy\ and \dzsq\ orbitals are reversed in reference \cite{glawion09}.  
In the following section we investigate the influence of magnetic order on the splitting in VOCl.

\subsubsection{Influence of the magnetic state}\label{sec:magstate}
\begin{figure}[hbt!]
\begin{center}
\includegraphics[scale=0.7]{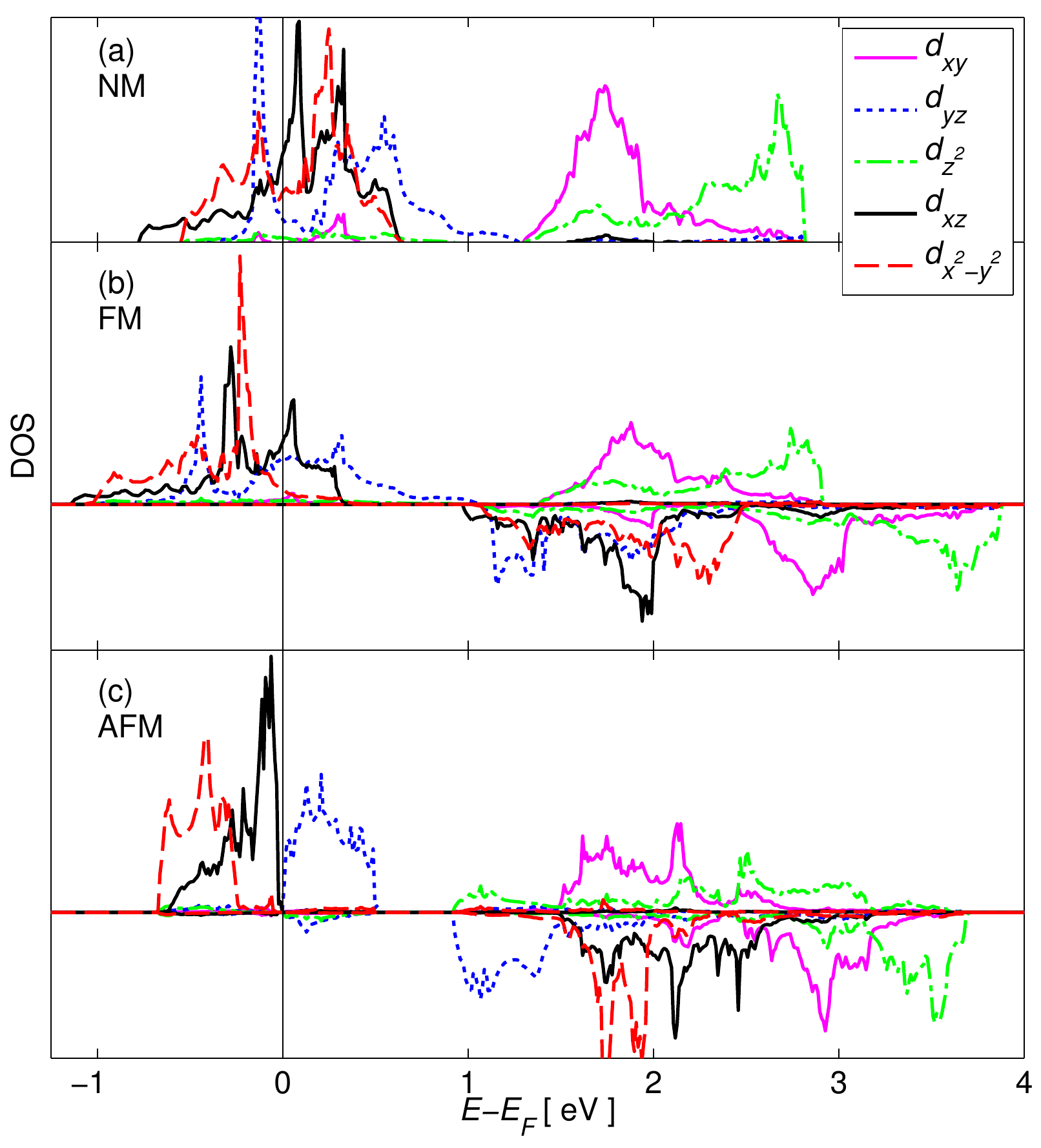}
\caption{DOS calculated with PBE at experimental lattice coordinates at 2~K for various magnetic orderings: (a) non-magnetic (b) ferromagnetic (c) antiferromagnetic.
\label{fig:pbe}
}
\end{center}
\end{figure}

Since our calculations have been performed with a different magnetic ordering as compared to those of previous work, it is interesting to consider the electronic spectrum for different models.
By switching off spin polarization altogether, we may reproduce the results presented reference \cite{glawion09} within 1~meV using plain PBE, i.e., $\U=0$. 
As shown in \fref{fig:pbe}(a), this also results in a metallic state, with partially occupied \dxsq, \dxz,\ and \dxy\, levels within 0.27 eV.
By allowing spin polarization and imposing FM order, the \dxsq\ and \dxy\ orbitals become split by 0.17~eV, although the \dxsq, \dxz, and \dxy\ orbitals are all still partially occupied, indicating a pronounced metallic regime (see \fref{fig:pbe}(b)).

\begin{figure}[hbt!]
\begin{center}
\includegraphics[scale=0.7]{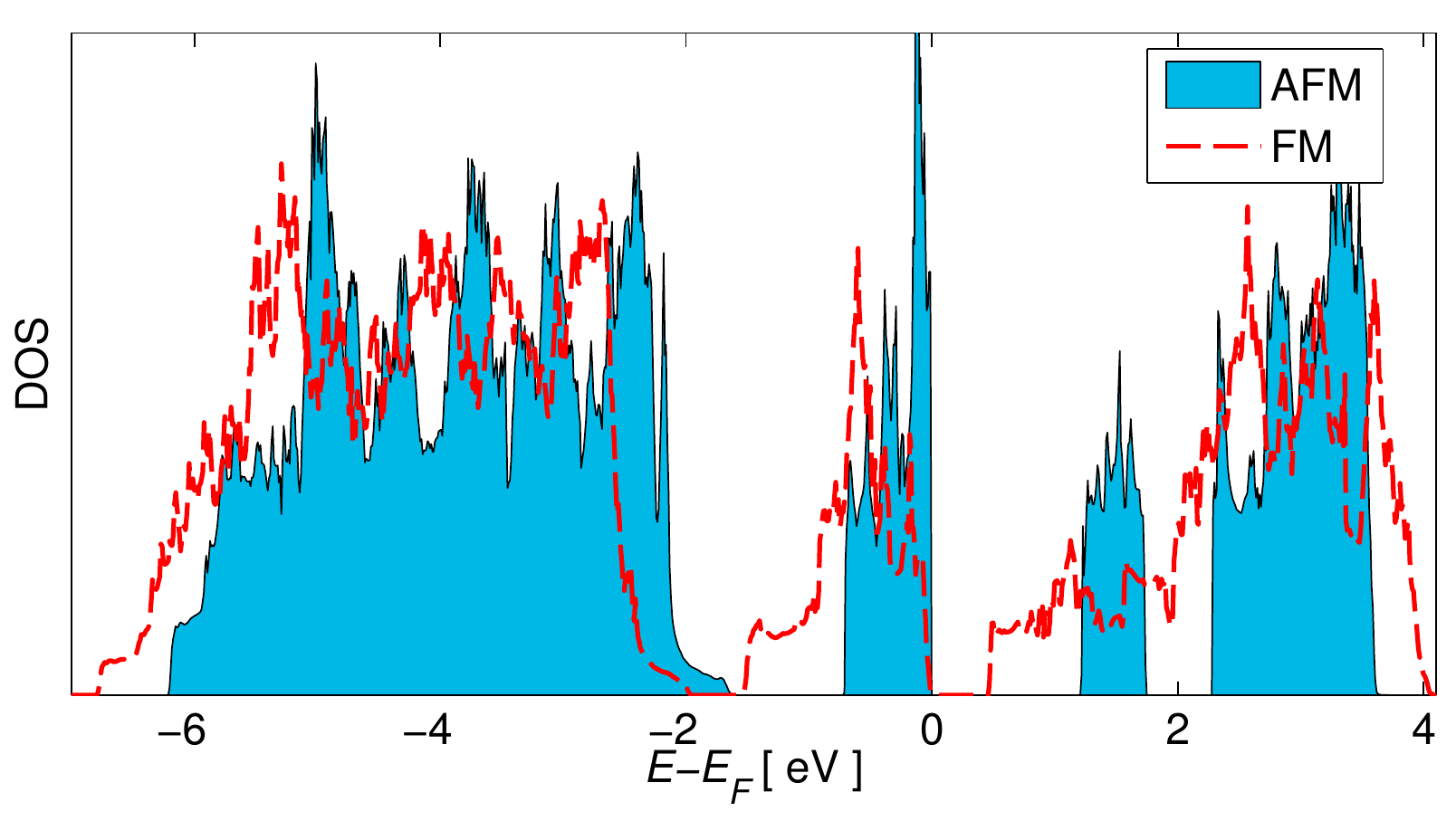}
\caption{\label{fig:FMvsAFM}
Total DOS calculated with FM and AFM order with $U=2$ eV.}
\end{center}
\end{figure}

Setting AFM order leads to depopulation of the \dyz\ level as the symmetry is broken, opening  a pseudogap at $E=E_F$ (see \fref{fig:pbe}(c)).
Using a nonzero $\U$ value will further increase the gap, and as seen in \fref{fig:FMvsAFM}, AFM order leads to much more narrow bands than FM order.
This translates to better agreement with experiment regarding the V-Cl/O and V-V charge gaps using AFM order.
In fact, with FM order the \dxsq\ and \dxy\ splitting is increased to 0.3 eV, which is significantly larger than experimental \cite{bogdanov11} results.

\subsection{Paramagnetic phase}\label{sec:results_orthorhombic}
In order to model magnetic disorder above the N\'eel temperature, we have constructed a supercell consisting of 108 atoms, where the local magnetic moments were distributed to mimic complete disorder in the thermodynamic limit.
Each V atom is then situated in a unique local environment, which in a sense is a mixture of the AFM and FM states.

With $\U=2$ eV, we obtain the volume 16.6 \ang, which is slightly larger than the experimental \cite{schafer61} room-temperature volume of 16.4 \ang.
The fully relaxed cell has orthorhombic symmetry, regardless of $\U$-value.
Switching to AFM order immediately leads to monoclinic distortion.
This shows that the observed monoclinic distortion is directly connected to AFM ordering of magnetic moments.

\begin{figure}[hbt!]
\begin{center}
\subfigure[\label{fig:SQStotal}]{
\includegraphics[scale=0.7]{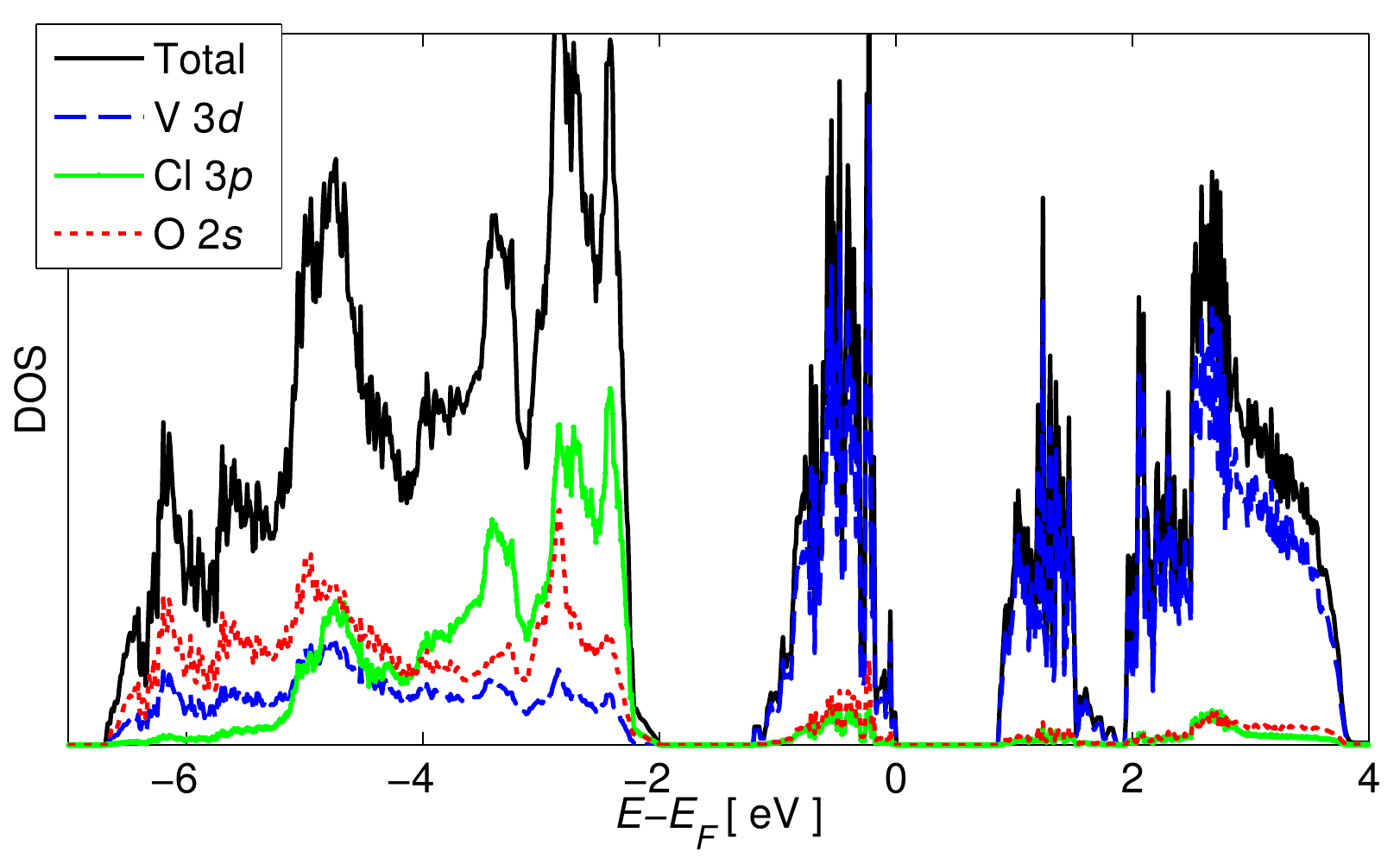}
}
\subfigure[\label{fig:SQSpartial}]{
\includegraphics[scale=0.7]{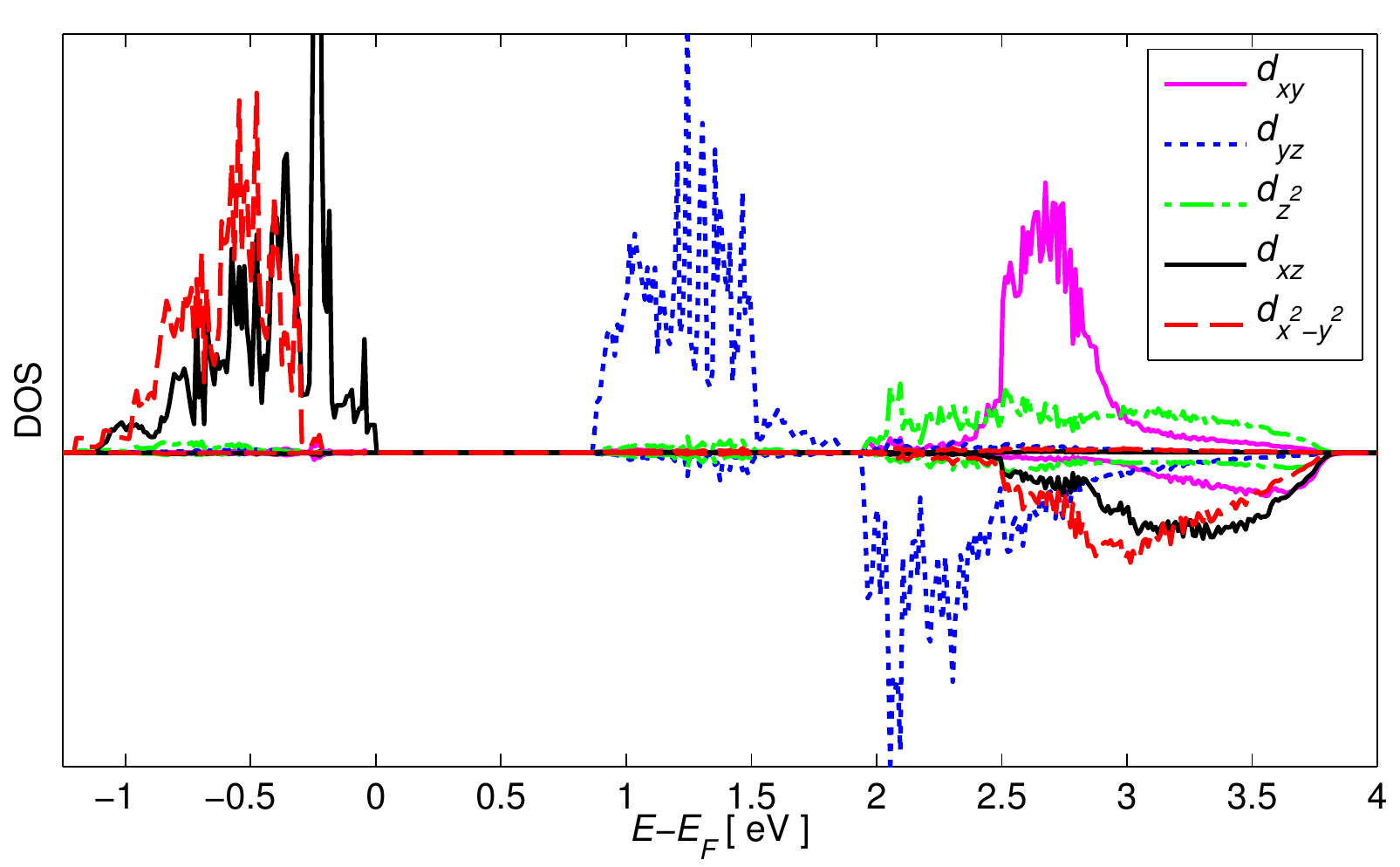}
}
\caption{\label{fig:SQSstatic} 
DOS for relaxed magnetically disordered 108-atom supercell.}
\end{center}
\end{figure}

\Fref{fig:SQSstatic} shows the DOS obtained with $\U=2$~eV.
The low-binding energy part is seen to be broadened due to the introduction of partial FM order, reducing the V-O/Cl and V-V gaps to 0.8 eV.
Averaging the orbital energies over all the V atoms in the supercell, we obtain the values presented in \tref{tab:enorb}.
The splitting between \dxsq\ and \dxz\ is seen to be very similar to the results for the AFM-ordered cell.

\begin{table}
\caption{\label{tab:enorb} Orbital energies with respect to the \dxsq\ level for various models, compared with the computational results presented in reference \cite{glawion09}.}
\begin{indented}
\item[]\begin{tabular}{l c | c c c c}
\br
             &      $\U$     &      \dxz &  \dyz  & \dzsq & \dxy    \\
                &    eV   &  eV &    eV &     eV &     eV  \\
                \mr
    AFM &  2.0 & 0.19  & 2.44 & 3.22 & 3.39 \\
             &  4.0  &  0.13   & 3.69   & 4.39  &  4.50  \\
\mr
SQS    &  2.0 & 0.18  &   2.47 & 3.34  & 3.46  \\
             &   4.0   & 0.12 & 3.69 & 4.47 & 4.54 \\
             \mr
  Reference \cite{glawion09} (theory) &    &  0.025   & 0.33  &  1.93  &   1.63 \\
    \br
 \end{tabular}
 \end{indented}
\end{table}

\section{Summary and conclusions}\label{sec:conclusions}
We have investigated the interplay between electronic, magnetic and structural degrees of freedom in VOCl by means of PBE+U calculations.
Our results show that the monoclinic distortion is a direct result of AFM order.
Disordered magnetic moments lead to an orthorhombic structure.
As a result of the distortion, we find the bond length indicated in \fref{fig:afm_order} to be shorter for AFM oriented pairs than of FM oriented pairs, supporting the magnetic structure proposed in references \cite{komarek09} and \cite{schonleber09}.

We also find that the \dxsq\ and \dxz\ levels are non-degenerate in both the paramagnetic and the magnetically ordered phase.
This puts VOCl on equal footing with TiOCl, where RIXS measurements have shown that these levels are non-degenerate \cite{glawion11}.
Our result on the orbital splitting is in contrast to the conclusions in reference \cite{glawion09}, but in line with the cluster calculations in reference \cite{bogdanov11} as well as experiments \cite{benckiser08}.
The discrepancy is related to the model for the magnetic state.
Assuming a paramagnetic state with vanishing local magnetic moments, levels are obtained as degenerate.
Including a finite local moment, the degeneracy is lifted.
Our work thus harmonizes PBE+U calculations with the calculations in reference \cite{bogdanov11}, which are explicit many-body calculations, but based on an embedded ferromagnetic atomic cluster.
However, we also find ferromagnetic calculations to yield strongly underestimated band gaps, as compared to experiment, due to the exaggerated band widths.

Our calculations do not include dynamical electrons correlations, which are crucial in TiOCl.
These can be included together with finite temperature electronic excitations using, e.g.,  dynamical mean field theory  \cite{georges96}.
Nevertheless, photoemission measurements in VOCl indicate that such fluctuations are less important than in TiOCl \cite{glawion09}.

Our work shows that further studies on VOCl and other transition metal oxychlorides imperatively need to take into account the relevant magnetic state. In particular, simulations for the paramagnetic phase must include disordered local magnetic moments.

\ack{M.\ E.\ is grateful to the Swedish e-Science Research Centre (SeRC) for financial support. All calculations were carried out using the facilities of the Swedish National Infrastructure of Computing (SNIC) at the National Supercomputer Centre (NSC), and the Centre for Scientific and Technical Computing (LUNARC).

The research of S.\ v.\ S.\ has been funded by the
Deutsche Forschungsgemeinschaft (DFG, German
Research Foundation) -- 386411512.}

\section*{References}
\bibliography{references}

\end{document}